\documentclass[twocolumn,showpacs,showkeys,preprintnumbers,amsmath,amssymb,prl]{revtex4}
\usepackage{graphicx}% Include figure files
\usepackage{dcolumn}% Align table columns on decimal point
\usepackage{bm}% bold math

\newcommand{\mcal}{\ensuremath{\mathcal}}
\newcommand{\opr}[1]{\ensuremath{\mathbf{\mathsf{#1}}}}

\newcommand{\expo}[1]{\ensuremath{\mbox{e}^{#1}}}
\newcommand{\abs}[1]{\ensuremath{\left|#1\right|}}

\newcommand{\kernelP}[2]{\ensuremath{\left<#1\left|#2\right|#1\right>}}

\begin{document}

%\preprint{wala pang preprint number!}

\title{Confined Quantum Time of Arrivals}% Force line breaks with \\

\author{Eric A. Galapon$^{1,2,3}$}
\email{eric.galapon@upd.edu.ph}
%\altaffiliation[Also at]{Physics Department, XYZ University.}%Lines break
% automatically
 %or can be forced with \\
\author{Roland F. Caballar$^1$}
\author{Ricardo T. Bahague Jr$^1$}%
\affiliation{$^1$Theoretical Physics Group, National Institute of Physics, University of
 the Philippines, Diliman, Quezon City, 1101 Philippines}
\affiliation{$^2$Theoretical Physics, The University of the Basque Country, Apdo. 
644, 48080 Bilbao, Spain}
\affiliation{$^3$Chemical Physics, The University of the Basque Country, Apdo. 
644, 48080 Bilbao, Spain}

\date{\today}% It is always \today, today,

\begin{abstract}
We show that formulating the quantum time of arrival problem in a segment of the real line suggests rephrasing the quantum time of arrival problem to finding states that evolve to unitarily collapse at a given point at a definite time. For the spatially confined particle, we show that the problem admits a solution in the form of an eigenvalue problem of a compact and self-adjoint time of arrival operator derived by a quantization of the classical time of arrival, which is canonically conjugate with the Hamiltonian in closed subspace of the Hilbert space.
\end{abstract}

%Use showkeys class option if keyword
\maketitle

The incorporation of time as a dynamical observable in quantum mechanics remains controversial \cite{gen}. And there is still no general consensus on how to do so, much less on what constitutes a meaningful quantum representation of time. This problem is widely known as the quantum-time-problem, and it takes on different facets.  One of these is the question whether time observables, such as time of arrivals, can be meaningfully represented by a self-adjoint operator or not, i.e. by a time operator. The basic requirement for a time operator is conjugacy with the Hamiltonian, in keeping with  quantum dynamics. However, the search for such an operator has been obstructed by nearly seven decades by Pauli's well-known theorem:  No such self-adjoint operator exists for semibounded Hamiltonians \cite{pauli,forpauli}. Thus it has been the consensus that one cannot introduce time operators without a compromise---Either self-adjointness or conjugacy but not both. The former, for example, has been advocated in constructing self-adjoint time of arrival operators \cite{self}. In recent years, the problem of introducing time in quantum mechanics has taken the latter route, which required time operators to be POVMs that transform covariantly under time translations \cite{povm}.

However, one of us has recently demonstrated that Pauli's theorem does not hold in a Hilbert space, and has proved the existence of self-adjoint characteristic time operators for semibounded Hamiltonians with compact inverses \cite{galapon}. These settle the question of the existence of self-adjoint operators canonically conjugate with a semibounded Hamiltonian. But addressing the existence issue has raised more issues. For example, these time operators may be canonically conjugate with the Hamiltonian in a closed subspace of the Hilbert space (a non-dense subspace, i.e. there exists a non-zero vector orthogonal to the subspace), and they can be bounded and compact, possessing a complete (normalizable) eigenfunctions with discrete (bounded) eigenvalues. The former property goes counter with the prevailing idea that a canonical pair must at least be canonical in a dense subspace to be meaningful. While the latter conflicts with the generally acknowledged basic covariance property of a meaningful time operator. The questions then arise as to what occasions these time operators appear and as to how we should interpret their eigenfunctions and eigenvalues.

To address these issues, we tackle in this Letter the free quantum time of arrival (QTOA) problem \cite{toa} anew. And we do so for two reasons. Firstly, the problem has been serving as the archetype of the perceived gross inability of standard quantum mechanics to accommodate the quantum aspects of time which has been traditionally attributed to Pauli's theorem. Secondly, the same problem provides us, ironically, with a perfect example of the existence of a self-adjoint, compact time operator, which is canonically conjugate with the Hamiltonian in a closed subspace. In this Letter, we show that formulating the QTOA-problem in a segment of the real line suggests rephrasing the problem to finding states that collapse at a given point, say at the origin, at a definite time. We show that, for a spatially confined particle, the problem admits a solution in the form of an eigenvalue problem of a compact and self-adjoint time of arrival operator derived by a quantization of the classical time of arrival.

The QTOA-problem is traditionally the problem of finding the time of arrival (TOA) distribution of a structureless particle prepared in some initial state. This operator is presumed to be the quantized classical-TOA in unbounded free space. That is, if a classical free particle, of mass $\mu$ in one dimension at location $q$ with momentum $p$, will arrive, say, at the origin at the time $T(q,p)=-\mu qp^{-1}$, then the quantum TOA-distribution must be derivable from the quantization of $T(q,p)$, from the operator $\opr{T}=-\mu (\opr{qp^{-1}+p^{-1}q})/{2}$. Formally the time of arrival operator $\opr{T}$ is canonically conjugate to the free Hamiltonian, $\opr{H}=(2\mu)^{-1}\opr{p}^2$, i.e. $[\opr{H},\opr{T}]=i\hbar$. It is well-known though that the resulting TOA-operator $\opr{T}$ is maximally symmetric and non-self-adjoint in unbounded space \cite{toa}. For a long time the non-self-adjointness of the quantized classical-TOA has been construed as a consequence of Pauli's theorem \cite{forpauli}.

First, let us show that the non-self-adjointness of $\opr{T}$ can be addressed by spatial confinement. Let the particle be confined between two points with length $2l$. This assumption is natural because all experiments are essentially bounded in space. If $p\neq 0$ and $\abs{q}<l$, the classical time of arrival at the origin (the first time of arrival, i.e. arrival without reflection from the boundaries) and the Hamiltonian are still  given by $T=-\mu qp^{-1}$ and $H=(2\mu)^{-1}p^2$, respectively; moreover, $T$ remains canonically conjugate with the  Hamiltonian. Then $\opr{T}$ is still the totally symmetric  quantized form of $T$ even when the particle is confined, and it likewise  remains canonically conjugate with the Hamiltonian.

To give meaning to $\opr{T}$, we attach the Hilbert space $\mcal{H}=L^2[-l,l]$ to the system. The position  operator is unique and is  given by the bounded operator $\opr{q}$,  $\left(\opr{q}\varphi\right)\!(q)=q\varphi(q)$ for all $\varphi(q)$ in $\mcal{H}$. On the other hand, the momentum operator and the Hamiltonian are not unique, and have to be considered carefully.  Our choice is dictated by the assumption of closedness of the system and the requirement of consistency with quantization: We assume the system to be conservative and we require that the evolution of the system be generated by a purely kinetic Hamiltonian. The former requires a self-adjoint Hamiltonian to ensure that time evolution is unitary. The later requires a self-adjoint momentum operator commuting with the Hamiltonian to ensure that the quantum Hamiltonian is the quantization of the purely kinetic Hamiltonian of the freely evolving classical particle between the boundaries. Now for every $\abs{\gamma}<\pi$, there exists a self-adjoint momentum operator given by  the operator $\opr{p_{\gamma}}=-i\hbar \partial_{q}$ whose domain consists of those vectors $\phi(q)$ in $\mcal{H}$ such that $\int\abs{\phi'(q)}^2\,dq<\infty$, satisfying the boundary condition  $\phi(-l)=\expo{-2i\gamma}\phi(l)$. With $\opr{p}_{\gamma}$ self-adjoint, the kinetic energy operator $\opr{K_{\gamma}}=\frac{1}{2\mu}\opr{p}_{\gamma}^2$ is consequently self-adjoint. Thus the Hamiltonian is purely kinetic, i.e. $\opr{H}_{\gamma}=\opr{K}_{\gamma}$.  The momentum and the Hamiltonian then commute and have the common set of plane wave eigenvectors.

Now let us consider $\opr{T}$ for $\gamma\neq 0$. Since  $\opr{q}$ appears in first power in $\opr{T}$, $\opr{T}$ is an operator if  the inverse of $\opr{p}_{\gamma}$ exists.  Since zero is not an eigenvalue of  $\opr{p}_{\gamma}$, the inverse $\opr{p_{\gamma}^{-1}}$ exists, and is in fact bounded and  self-adjoint. Then it follows that, for every $\abs{\gamma}<\pi$, $\opr{T}$ is a  bounded, symmetric operator. Thus $\opr{T}$ is self-adjoint. For a given $\gamma$, we identify $\opr{T}$ with the operator $\opr{T}_{\gamma}=-\mu(\opr{q}\opr{p_{\gamma}^{-1}}+\opr{p_{\gamma}^{-1}}\opr{q})2^{-1}$ derived from the formal operator $\opr{T}$ by replacing $\opr{p}$ with $\opr{p}_{\gamma}$. We shall refer to $\opr{T}_{\gamma}$ as the non-periodic confined time of arrival (CTOA) operator for a given $\abs{\gamma}<\pi$. In coordinate representation, $\opr{T}_{\gamma}$ becomes the Fredholm integral operator $\left(\opr{T}_{\gamma}\varphi\right)\!(q)=\int_{-l}^{l}T_{\gamma}(q,q')\,\varphi(q')\,dq
',$  for all $\varphi(q)$ in $\mcal{H}$, where the kernel is given by 
\begin{equation} \label{repre}
T_{\gamma}(q,q')=-\mu\frac{(q+q')}{4\hbar\sin\gamma}\left(e^{i\gamma}
\mbox{H}(q-q')+e^{-i\gamma}\mbox{H}(q'-q) \right),
\end{equation}
in which H$(q-q')$ is the Heaviside function. 

For $\gamma=0$, we face a different problem. Because zero is now an eigenvalue  of the momentum operator $\opr{p_0}$, $\opr{p_0}^{-1}$  does not  exist, and operator $\opr{T}$ is meaningless. But this pathology can be rigorously treated by projecting $\opr{p}_0$ onto the subspace orthogonal to its null subspace, as discussed in \cite{galapon}. Following \cite{galapon}, $\opr{T}$ corresponds to a compact, self-adjoint integral operator $\opr{T}_0$ whose kernel is
\begin{equation} \label{periodic}
	T_{0}(q,q')=\frac{\mu}{4i \, \hbar}(q+q')\mbox{sgn}(q-q')-\frac{\mu}{4i\,\hbar
 l}\left(q^2 -q'^2\right).
\end{equation}
We shall refer to this as the periodic CTOA operator. 

With the above representation, one can show that $\opr{H}_{\gamma}$ and $\opr{T}_{\gamma}$ form a canonical pair in a closed subspace of $\mcal{H}$ for every $\gamma$. Moreover, the kernel $T_{\gamma}(q,q')$ of $\opr{T}_{\gamma}$ is square integrable, i.e. $\int_{-l}^l\int_{-l}^l\abs{T_{\gamma}(q,q')}^2\, dq dq'<\infty$. This means that $\opr{T}_{\gamma}$ is compact: it has a complete set of (square integrable) eigenfunctions and its spectrum is discrete.

For the $\gamma\neq 0,\frac{\pi}{2}$ CTOA-operators, it can be shown that to every positive integer $n=1, 2, \dots$ there is a pair of eigenfunctions $\varphi_{n,\gamma}^{\pm}(q)$, with equal magnitudes of eigenvalues and of opposite signs, i.e. $\tau_{n,\gamma}^{+}=-\tau_{n,\gamma}^{-}$; the sign indicates the sign of the eigenvalue. The number $n$ corresponds to the $n$-th positive root, $r_n$, of $J_{-\frac{3}{4}}(x)J_{-\frac{1}{4}}(x)-\cot^2\gamma J_{\frac{3}{4}}(x)J_{\frac{1}{4}}(x)=0$, where $J_{\nu}(x)$ is the Bessel function of the first kind. The eigenfunctions are given by
\begin{eqnarray}
\varphi_{n,\gamma}^{\pm}\!\!\!\!& &\!\!\!\!(q)\!\!=\!\!A_n e^{\mp ir_n\frac{q^2}{l^2 }}\!\! \left[ J_{\frac{3}{4},\frac{1}{4}}^{\mp}\!\!\left( r_n\frac{q^2}{l^2}\right)\!\left(J_{-\frac{1}{4}}\!(r_n)\!-\!\cot\gamma J_{\frac{3}{4}}\!(r_n)\right)\right.\nonumber\\
& &\left.\!\! \pm \frac{2 q \sqrt{r_n}}{l}J_{\frac{1}{4},\frac{3}{4}}^{\mp}\!\!\left( r_n\frac{q^2}{l^2}\right)\! \left({J_{-\frac{3}{4}}\!(r_n)-\cot\gamma J_{\frac{1}{4}}\!(r_n)}\right)\right],\label{xxx}
\end{eqnarray}
where $J_{\nu,\rho}^{\mp}(x)=x^{\nu}(J_{-\nu}(x)\mp i J_{\rho}(x))$, and $A_n$ is the normalization constant. The corresponding eigenvalues are $\tau_{n,\gamma}^{\pm}=\pm (\mu l^2/4\hbar)\,r_n^{-1}$. We shall call those that do not vanish anywhere in $[-l,+l]$ as nonnodal-eigenfunctions; otherwise, as nodal eigenfunctions. The non-nodal (nodal) eigenfunctions correspond to those with even (odd) quantum number $n$. A nodal is zero at some single point $q_0$.

%%%%%%%%%%%%%%%%%%%%Figure 1%%%%%%%%%%%%%%%%%%%%%%%%%%%%%%%
\begin{figure}[!tbp]
{\includegraphics[height=1in,width=3in,angle=0]{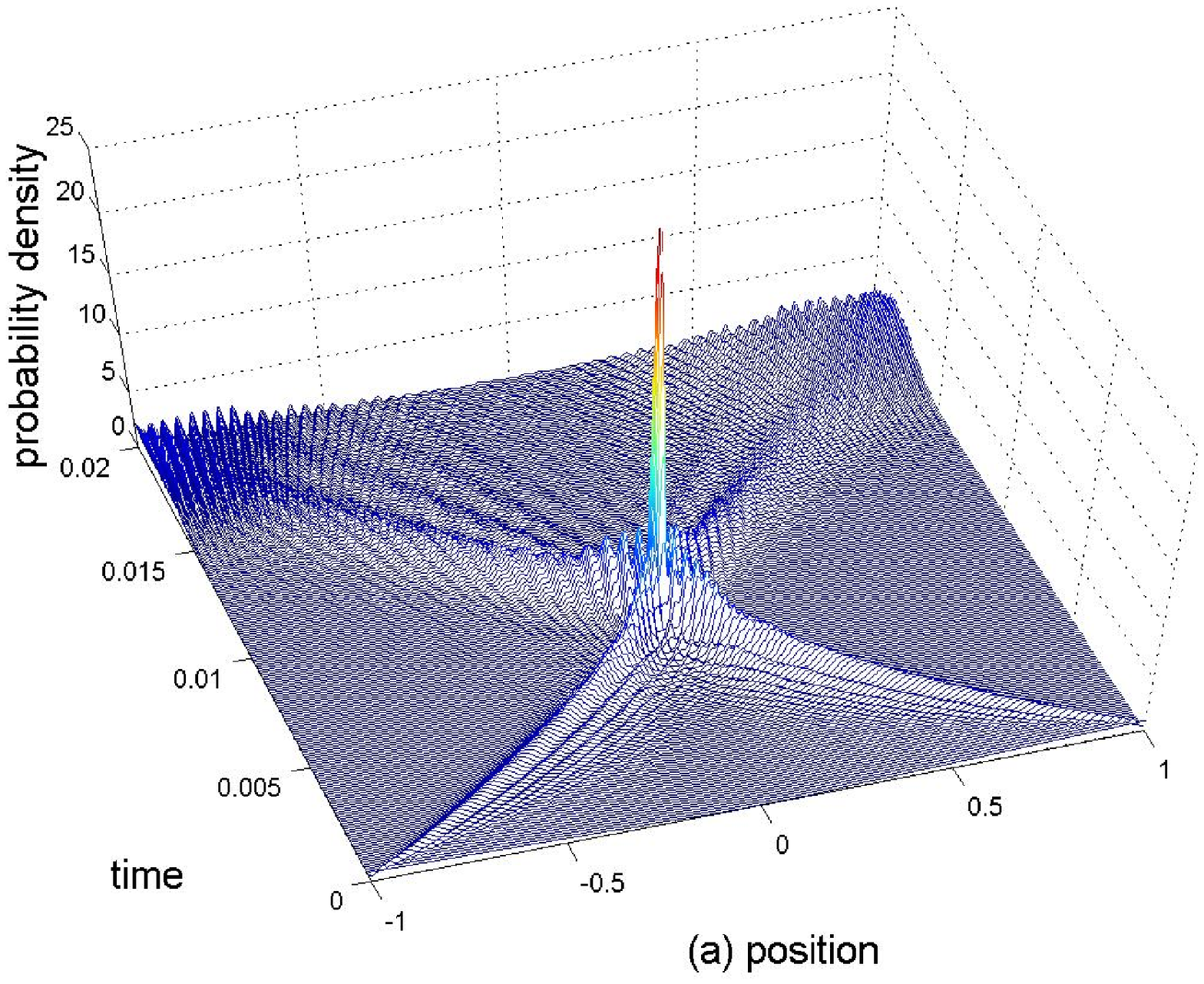}}
{\includegraphics[height=1in,width=3in,angle=0]{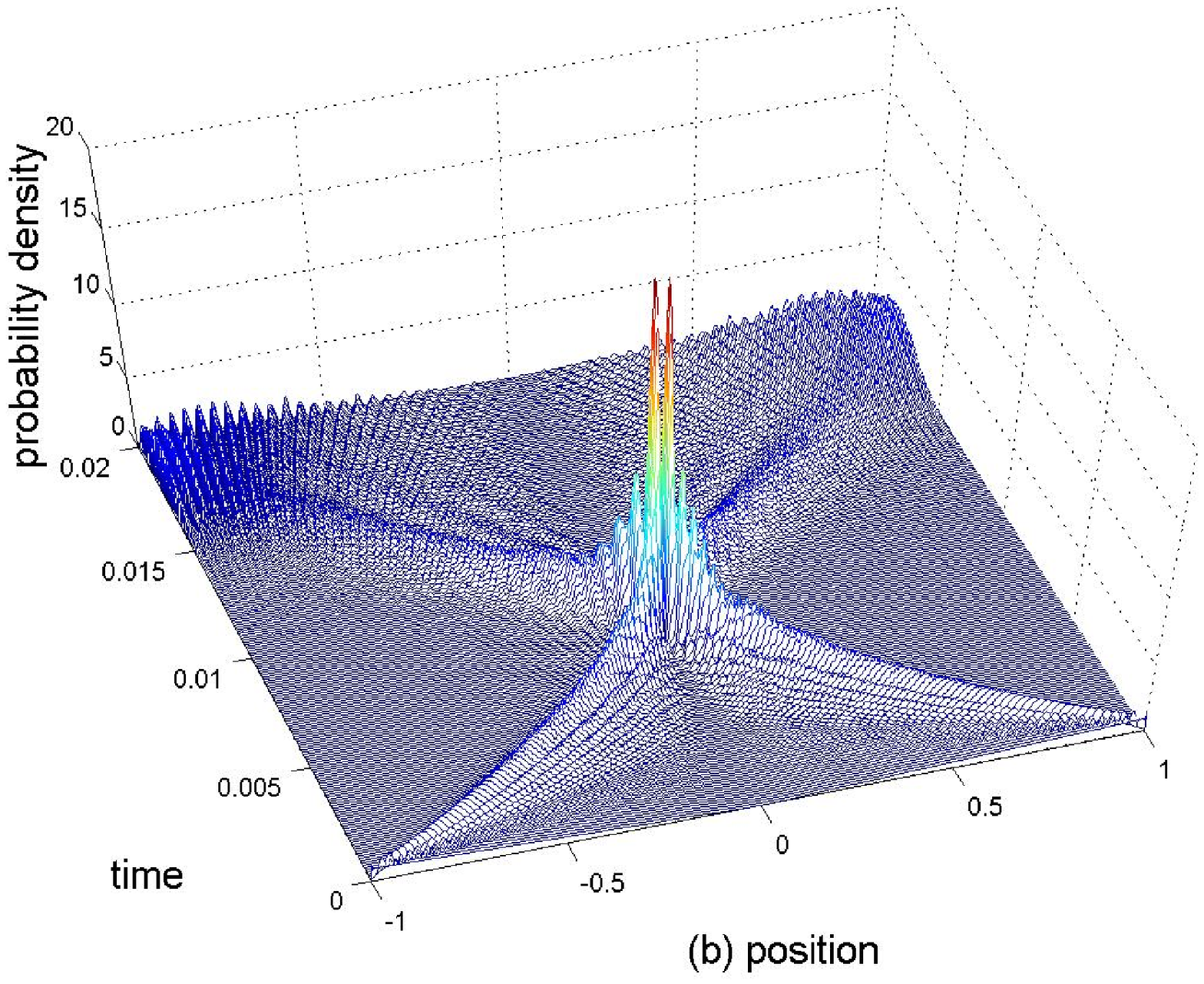}}
\caption{The $n=20$ ({\it a}) and $n=21$ ({\it b}) evolved probability densities for $\gamma=0.01$, with $\hbar=l=m=1$. Both unitarilly collapse at the origin at their respective eigenvalues, 0.0081 and 0.0079.}
\end{figure}
%%%%%%%%%%%%%%%%%%%%%%%%%%%%%%%%%%%%%%%%%%%%%%%%%%%%%%%%%%%%%%%%

The eigenfunctions for $\opr{T}_{\frac{\pi}{2}}$ can be derived directly from equation-({\ref{xxx}). In this case, the eigenfunctions bifurcate into odd and even eigenfunctions. The even and non-nodal eigenfunctions are $\varphi_{s,\frac{\pi}{2},e}^{\pm}(q)=\varphi_{s,\frac{\pi}{2}}^{\pm}(q)$,
with the eigenvalues given by $\tau_{s,\frac{\pi}{2},e}^{\pm}=\pm(\mu l^2/4\hbar)r_s^{-1}$, and the $r_s$'s are the positive roots of $J_{-\frac{3}{4}}(x)=0$, with $s=1, 2, \dots$. And the odd and nodal eigenfunctions are  $\varphi_{u,\frac{\pi}{2},o}^{\pm}(q)=\varphi_{u,\frac{\pi}{2}}^{\pm}(q)$, with the eigenvalues given by $\tau^{\pm}_{u,\frac{\pi}{2},o}=\pm (\mu l^2/4\hbar)r_u^{-1}$, and the $r_u$'s are the positive roots of $J_{-\frac{1}{4}}(x)=0$, with $u=1,2,\dots$. 

The eigenfunctions of $\opr{T}_{0}$ are likewise either even or odd. The even and non-nodal eigenfunctions are given by
\begin{equation}
\varphi_{s,0,e}^{\pm}(q)=B_s  e^{\mp i\frac{q^2}{l^2} r_s}\,J_{\frac{3}{4},\frac{1}{4}}^{\mp}\!\!\left(\frac{q^2}{l^2} r_s\right) +  \frac{4 B_s e^{\mp ir_s} J_{\frac{1}{4}}(r_s)}{(4 r_s)^{\frac{1}{4}}}\nonumber
\end{equation}
where $B_s$ is the normalization constant. The corresponding eigenvalues are $\tau_{n,0}^{\pm}=\pm (\mu l^2/4\hbar)r_s^{-1}$, where the $r_s$'s are roots of the equation $J_{-\frac{3}{4}}(x)+\frac{2}{3}J_{\frac{5}{4}}(x)+\frac{1}{x}J_{\frac{1}{4}}(x)=0$, with $s=1, 2, \dots$. The odd and nodal eigenfunctions, $\varphi_{u,0,o}^{\pm}(q)$, together with their corresponding eigenvalues, coincide with those of $\varphi_{u,\frac{\pi}{2},o}^{\pm}(q)$.

Now we turn to the dynamics of the above eigenfunctions. Using symmetry arguments, it can be established that the negative eigenvalue-eigenfunctions have exactly the same dynamics as those of the positive eigenvalue-eigenfunctions in the time reversed direction. It is then sufficient for us to consider in detail the dynamical behaviors of the positive eigenvalue eigenfunctions. Our analysis is based on the numerical evaluation of the evolution law $\varphi(t)=e^{-i\opr{H}_{\gamma}t/\hbar} \varphi(0)$ in energy representation. 

Figure-1.{\it a} shows the general features of the non-nodal eigenfunctions for every $\gamma$. The probability density $\abs{\varphi_{n,\gamma}^+(q,t)}^2$ collapses with non-vanishing width at the origin, obtaining its maximum value there at the time equal to the eigenvalue $\tau_{n,\gamma}^+$ (within numerical accuracy). Moreover, the variance of the position operator in the eigenfunction $\varphi_{n,\gamma}^+(q,t)$, $\sigma^2_{\varphi_{n,\gamma}}(t)=\kernelP{\varphi^+_{n,\gamma}(t)}{\opr{q}^2}-\kernelP{\varphi^+_{n,\gamma}(t)}{\opr{q}}^2$, is minimum at the eigenvalue $\tau_{n,\gamma}^{+}$. On the other hand, Figure-1.{\it b} shows the general features of an evolving nodal eigenfunction. The probability density of a nodal eigenfunction collapses towards the origin, its zero being at the orgin at the eigenvalue, i.e. $\abs{\varphi_{n,\gamma}^+(0,\tau_{n,\gamma}^+)}^2=0$. In contrast to non-nodals with one peak at the origin, nodals evolve to have two peaks approaching the origin. In both cases, the minimum variance of each eigenfunction asymptotically decreases as $n^{-v}$ for some $1<v<2$. The eigenfunctions then become arbitrarily localized at the origin at their eigenvalues for arbitrarily large $n$.

%%%%%%%%%%%%%%%%%%%%Figure 1%%%%%%%%%%%%%%%%%%%%%%%%%%%%%%%
\begin{figure}[!tbp]
{\includegraphics[height=1.1in,width=3.25in,angle=0]{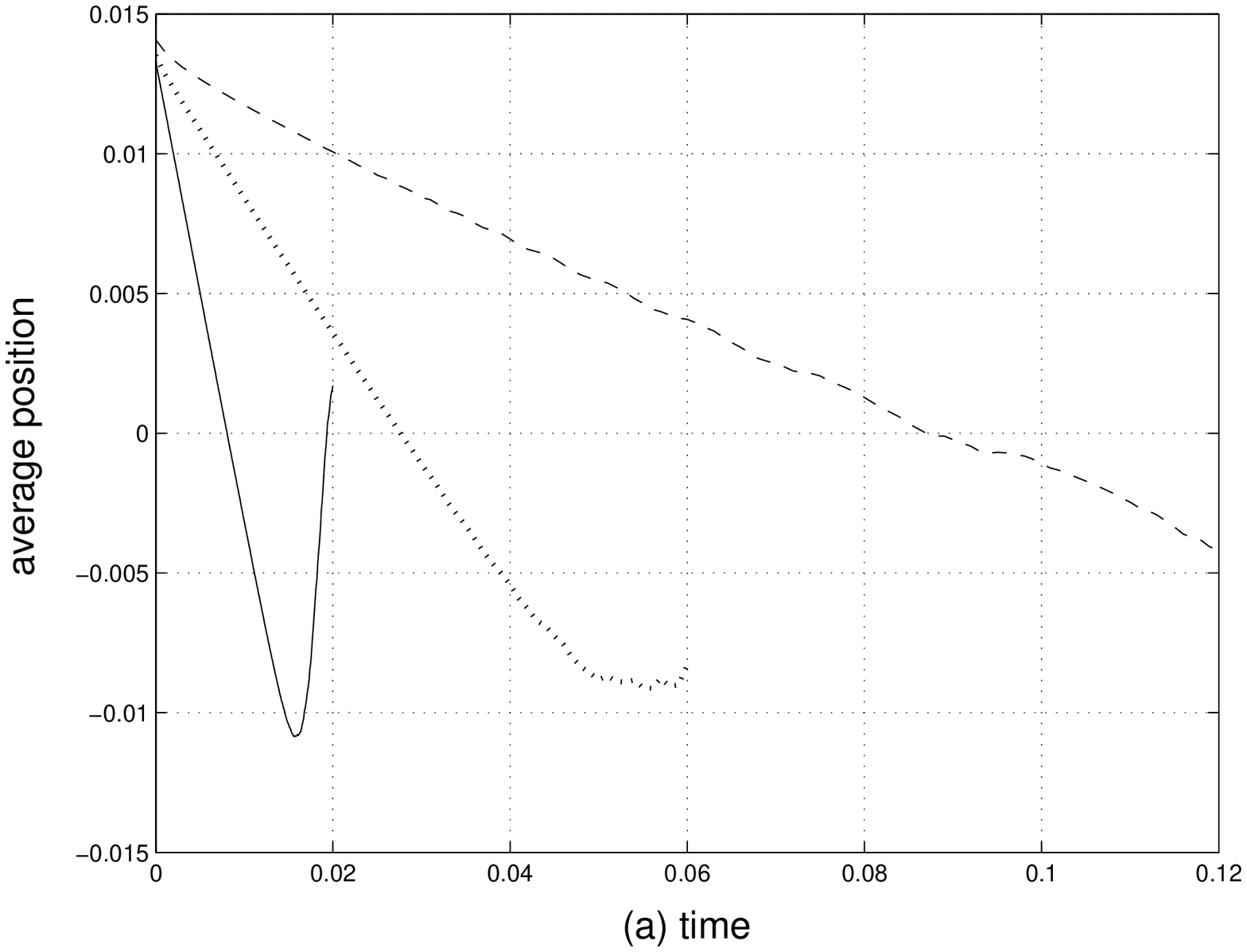}}
{\includegraphics[height=1.1in,width=3.25in,angle=0]{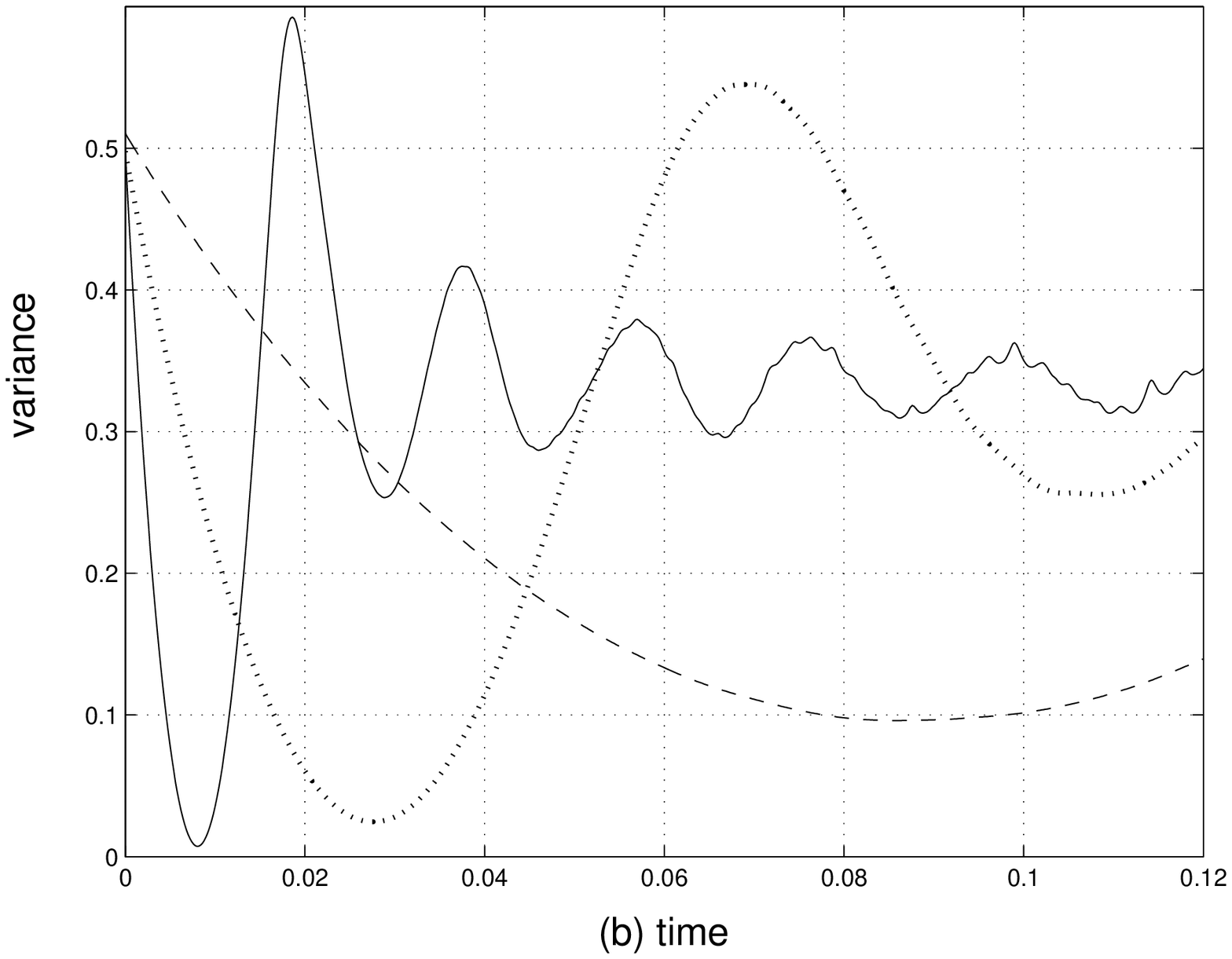}}
\caption{Figure-{\it a} shows the expectation value of the position operator for $n=2$ (dashed line), $n=6$ (dotted line), and $n=20$ (solid line) eigenfunctions. They cross the origin at the respective eigenvalues of the eigenfunctions, 0.0899, 0.0276, and 0.0081. Figure-{\it b} shows the variances, $\sigma^2(t)$, for the same eigenfunctions as a function of time. Their variances obtain their minimum values at their respective eigenvalues. The parameters are the same as in Figure-1.}
\end{figure}
%%%%%%%%%%%%%%%%%%%%%%%%%%%%%%%%%%%%%%%%%%%%%%%%%%%%%%%%%%%%%%%%

The eigenfunctions for $\gamma\neq 0, \frac{\pi}{2}$, non-nodal or nodal, are non-parity eigenfunctions so that they exhibit assymetry in their probability densities. For these eigenfunctions, the expectation value of the position operator with respect to them, i.e. $\left<\opr{q}(t)\right>_{\varphi_{n,\gamma}}=\kernelP{\varphi^+_{n,\gamma}(t)}{\opr{q}}$, follows the classical trajectory, at least in the time interval $0\leq t\leq\tau_{n,\gamma}^+$,  and assumes the value zero at the eigenvalue $\tau_{n,\gamma}^+$ \cite{ex}. It can be established that these non-parity eigenfunctions are the quantum analogues of the classical case where the initial position and momentum are non-vanishing. Figure-2 shows the general features of the position expectation value and variance as a function of time for non-parity eigenfunctions. On the other hand, the expectation value of the position and momentum operators in the eigenfunctions for $\gamma=0,\frac{\pi}{2}$vanishes. This shows that the $\opr{T}_{\gamma=0,\frac{\pi}{2}}$ CTOA-operators are quantizations of the classically indeterminate case where the initial position and momentum are vanishing.

Our numerical results then strongly support the following interpretation for the confined-TOA eigenfunctions and their corresponding eigenvalues: A CTOA eigenfunction is a state that evolves to unitarily collapse or to arrive at the origin at its eigenvalue along a classical trajectory, i.e. a state in which the events of the centroid being at the origin and the position distribution width being minimum occur at the same instant of time equal to the eigenvalue, with the centroid following the classical trajectory. But there is one thing more fundamental about these eigenfunctions---they are purely geometrical. The parameters of the eigenfunctions are only $l$ and $\gamma$, which are geometrical in origin: $l$ determines the volume of the configuration space; $\gamma$, the translation property. In particular, they do not depend on $\hbar$ and on the mass $\mu$ of the particle, only the eigenvalues depend on them. This means that the CTOA-eigenfunctions are universal, applying to all massive particles. On first thought this is unexpected, but it may in fact be a physical necessity. In classical mechanics, the initial position of the particle can be chosen independent of its mass; the choice of the initial position is only dictated by the geometry of the configuration space, i.e. where the particle is only possible to place. And so may be in quantum mechanics. This is reflected in the pure geometric nature of the CTOA-eigenfunctions that assumes the initial ``position'' of the quantum particle, which is everywhere at once in the entire length $2l$. We point out that the eigenfunctions and eigenvalues of the CTOA-operators are tied with the dynamics of the system; hence, they acquire interpretation independent from the quantum measurement postulate. Moreover, the dicreteness and boundedness of the eigenvalues indicate that covariance, which implies the eigenvalues of a time operator to be the entire real line, is not necessary for a time operator to be meaningful.

Our results suggest introducing the concept of ideal time of arrival states in a given Hilbert $\mathcal{H}_{\Omega}$ over some configuration space $\Omega$, and rephrasing the time of arrival problem in terms of these states. For a fixed initial time $t_0=0$, let us define the subspace, $\mathcal{H}_x$, of $\mathcal{H}_{\Omega}$ characterized by the point $x$ interior to $\Omega$. A vector $\varphi$ in $\mathcal{H}_{\Omega}$ is in $\mcal{H}_x$ if ({\it a}) there exists a time $\tau>0$ such that $\sigma^2_{\varphi}(\tau)<\sigma^2_{\varphi}(t)$ for all $t\neq\tau$; if ({\it b}) $\left<\opr{q}(\tau)\right>_{\varphi}=x$; if ({\it c}) $\left<\opr{q}(t)\right>_{\varphi}$ follows the classical trajectory at least within the time interval $0\leq t\leq \tau$; and if ({\it d}) $\varphi$ is purely geometrical. We refer to these states as the ideal time of arrival states of the system $\mathcal{H}_{\Omega}$ with respect to the driving Hamiltonian $\opr{H}$, and the associated times $\{\tau\}$ as their respective times of arrivals at the point $x$. If there is perfect recurrence, then we refer to $\tau$ as the first arrival time. Now we can pose the problem: Given the system $\mcal{H}_{\Omega}$ driven by the Hamiltonian $\opr{H}$ and given a point $x$ interior to $\Omega$, what are the ideal time of arrival states $\mcal{H}_x$ and their corresponding arrival times? Because the wavefunction has a definite trajectory dictated by the Schr{\"o}}dinger equation in the Hilbert space, this is a well defined quantum mechanical problem.

And this problem may or may not have a solution depending on the given $\mathcal{H}_{\Omega}$ and $\opr{H}$. The confined time of arrival problem we have considered here is a benchmark problem having a solution in the form of the standard quantum mechanical eigenvalue problem of the quantized classical time of arrival. However, the solution we have obtained here need not be the only solution because other means of quantizing the classical time of arrival may lead to a different set of solutions. On the other hand, the unconfined time of arrival problem is a benchmark problem that does not posses a solution in the form of an eigenvalue problem of the quantized classical time of arrival---the eigenfunctions are non-square integrable and non-geometrical because of their dependence on the mass of the particle \cite{toa}. It is of fundamental significance to understand why these two related problems have divergent characteristics. An understanding may lead us to a deeper insight into the quantum time problem, and into the foundations of quantum mechanics itself. The confined quantum time of arrival problem has already tought us one lesson---Quantum mechanics is not at all inept at addressing the quantum aspects of time if only we knew what question to ask.

This work has been supported by the National Research Council of the Philippines  through grant number I-81-NRCP, and partially supported by the ``Ministerio de Ciencia y Tecnolog{\'\i}a'' and FEDER (Grant BFM2003-01003). EAG is by the University of the Philippines System through the U.P. Creative and Research Scholarship Program. This paper has benefited from discussions with A. M. Steinberg, S. Jhingan, R. de la Madrid, I. Egusquiza, and J.G. Muga.


\begin{thebibliography}{00}
\bibitem{ex} Our numerical method failed to converge up to the eigenvalues of the largest eigenvalue-eigenfunctions for $\gamma\neq 0,\frac{\pi}{2}$. We expect though that they should behave dynamically as the rest of the eigenfunctions.
\end{thebibliography}
\end{document}